\documentclass[prd,twocolumn,floats,floatfix,showpacs,nofootinbib]{revtex4}
\usepackage{graphicx}
\usepackage{graphics}
\usepackage{amssymb}
\usepackage{amsmath}
\usepackage{url}
\usepackage{color}
\newcommand{\fracc}[2]{\frac{\textstyle{#1}}{\textstyle{#2}}}

\newcommand{\p}{\partial}
\begin{document}

\title{On the effective acoustic geometry for relativistic viscous fluids}

\author{E. Bittencourt}\email{bittencourt@unifei.edu.br}
\author{V. A. De Lorenci}\email{delorenci@unifei.edu.br}
\author{R. Klippert}\email{klippert@unifei.edu.br}
\author{L. S. Ruiz}\email{lucasruiz@unifei.edu.br}

\affiliation{Universidade Federal de Itajub\' a, Av.\ BPS 1303, Itajub\' a, Minas Gerais 37500-903, Brazil}

\date{\today}

\begin{abstract}
Hadamard-Papapetrou method of field discontinuities is here employed in order to determine the effective metric that describes the propagation of acoustic perturbations in isentropic fluids. It is shown that, when dissipative effects are present, small perturbations in fastly moving fluids have a natural description in terms of an effective acoustic geometry. As an application of our results, a model for an acoustic black hole in viscous fluid is investigated.
\end{abstract}

\maketitle

\section{Introduction}

Analog models for general relativity have been considered in the literature for many years. The first work exploring this idea was published almost a century ago \cite{gordon1923} in the context of optics in moving dielectrics. Following this seminal work some related results were reported, most of them based on electrodynamics (see for instance Refs.\ \cite{plebanski1960,felice1971,schenberg1971}). In the realm of hydrodynamics, a sonic analog of a black hole geometry was proposed in 1981 \cite{unruh1981}, motivating the search for systems exhibiting an analog event horizon that could be experimentally realizable. One important issue in developing analog models of a black hole is the expectation of measuring analog Hawking radiation in laboratory. Despite of the predicted very low temperature associated with such phenomenon, techniques based on laser cooling and its application in measurements in Bose-Einstein condensates led to new achievements \cite{garay2000} (see Ref.\ \cite{blv} for a review) in this research area. Experimental evidences of the existence of Hawking radiation in analog systems were already reported \cite{steinhauer2014,steinhauer2016}.

The evolution of small perturbations in the limit of geometrical optics is commonly achieved by using the eikonal approximation, the standard procedure of which is to derive the dispersion relation from the dynamical equations for the perturbations. In the context of nonlinear electromagnetic theories, for instance, the dispersion relation is defined not only in terms of the background metric, but it also involves the unperturbed values of the fields of the theory, favoring the definition of an emergent effective metric \cite{nov_vis_volo}. In fluid mechanics this is not different, as one can see in Ref.\ \cite{visser,visser_molina}, where the deduction of the effective metric demands a quite long calculation using the eikonal approximation even for ideal fluids. An immediate comparison shows that the effective metric is much less manageable in the acoustic case than in the electromagnetic case due to the number of free parameters encoded in each metric tensor. This could be seen as an issue for those who want to propose some analog model for gravity in acoustic configurations, since the flow velocity may require large values in order to achieve the desired behavior, in confront with the regime of applicability of the geometric description \cite{vit_obukov}.

If one's interest points directly to the analogy between the geometrical description of perturbations within a medium in the context of analog models, then the equations of motion for the perturbations are not necessary. It is enough to obtain the dispersion relation associated to the perturbations, from which one can directly read the effective metric. With this in mind, it is worth noticing that there are equivalent methods to derive the dispersion relations which are more appealing than the eikonal expansion for the analysis of kinematical aspects of the fluid perturbations. This is the case of the Hadamard-Papapetrou \cite{hadamard,papapetrou} method, used in the study of shock waves, which deals with computing discontinuities of the derivative of continuous fields across a moving hypersurface. Although this method fails when discontinuous fields are considered, a hybrid technique illustrated in Ref.\ \cite{dante} deals properly with such cases.

In this paper, we shall apply the Hadamard-Papapetrou method in the context of fluid dynamics in order to derive the effective metric for a given fluid. In opposition to the eikonal approach, this method does not lead to a unique effective metric, but to a whole class of such metrics that are conformally related to one another. A suitable choice of the conformal factor makes the results obtained here to agree with previous calculations.

To the best of our knowledge such method has never been explored for fluids before (see \cite{nov_vis_volo,blv} and references therein), in despite of the fact that it is widely used in the context of geometric optics. Our aim here is to scrutinize its consequences for the geometrical description considering a larger class of fluids. In particular, we easily obtain an effective metric for barotropic perfect fluids, and generalize the results to the case of dissipative fluids.

The following section provides the general setup of shock waves for the geometric acoustics of isentropic fluids in the non-relativistic regime. Sec.~\ref{II} deals with the full relativistic case. The standard effective geometry for the acoustic perturbations of a perfect fluid is described by means of shock waves in Sec.~\ref{perfect}, and the generalization for dissipative systems is given in Sec.~\ref{viscous}. As an application of the method, the acoustic analog model of a black hole for viscous fluids is provided in Sec.~\ref{bh}. Summary and concluding remarks are presented in Sec.~\ref{V}.

The speed of light in vacuum is set to unit. We use the Lorentz signature $[+,-,-,-]$. Latin indices $i,j,k\ldots$ run from 1 to 3 (the three spatial directions) and repeated indices in a monomial implicitly indicate summation. Notice that up and down indices are not distinguished in Section~\ref{propagation}.

\section{Wave propagation in geometric acoustics}
\label{propagation}
A non-relativistic ideal fluid in thermodynamical equilibrium is characterized by its mass density $\rho$, pressure $p$ and flow velocity $v_i$ with a given equation of state $p=p(\rho)$, its dynamics being completely determined by the continuity and the Euler equations \cite{landau}. If these equations are to be responsible for the evolution of the fluid in the presence of viscosity as well, however, the dissipative components should be phenomenologically set in terms of the kinematical properties of the fluid. In particular, the continuity equation should not change in the presence of dissipative terms
\begin{equation}
\label{nr_cont}
\frac{\p\rho}{\p t}+\frac{\p}{\p x_i}(\rho v_i)=0,
\end{equation}
while the Euler equation ought to be modified. For later convenience, we shall rewrite it in a compact form as
\begin{equation}
\label{nr_euler}
\frac{\p}{\p t} (\rho v_i)+\frac{\p}{\p x_j}\Pi_{ij}=0,
\end{equation}
where $\Pi_{ij}$ is the tensor of density of flux of momentum. For viscous fluids, we define
$$\Pi_{ij}=p\delta_{ij}+\rho v_i v_j -\pi_{ij},$$
where $\pi_{ij}$ is the viscous stress tensor. For Newtonian fluids, the most general expression of \(\pi_{ij}\) is
$$\pi_{ij}=\eta\left(\frac{\p v_i}{\p x_j} + \frac{\p v_j}{\p x_i} -\frac{2}{3}\frac{\p v_l}{\p x_l}\delta_{ij} \right)+\xi\delta_{ij}\frac{\p v_l}{\p x_l},$$
with $\eta$ and $\xi$ as the shear and bulk viscosity constant coefficients, respectively, and $\delta_{ij}$ denotes the 3-dimensional Kronecker delta.

For instance, in the simple case of a motionless fluid, the wave equation for the perturbations can be easily obtained by assuming $\rho=\rho_0+\rho_1$ and $p=p_0+p_1$, where $\rho_0$ and $p_0$ are constants corresponding to the background and $\rho_1$ and $p_1$ are small perturbations of those quantities. This leads to the equation
\begin{equation}
\frac{\p^2\rho_1}{\p t^2} - c_s^2\frac{\p^2\rho_1}{\p x_i\p x_i} - \frac{1}{\rho_0}\left(\frac{4}{3}\eta+\xi\right)\frac{\p}{\p t}\left(\frac{\p^2\rho_1}{\p x_i\p x_i}\right)=0
\label{d3},
\end{equation}
where
\begin{equation}
c_s:=\sqrt{p'}=\left.\sqrt{\dfrac{\partial p}{\partial\rho}}\right|_{{p=p_0},{\rho=\rho_0}}
\label{soundbkg}
\end{equation}
is the speed of sound in the fluid. Equation (\ref{d3}) cannot be expressed in terms of the wave operator solely due to the presence of the dissipative term. The eikonal expansion for the solutions of this equation leads, at first order of approximation, to the dispersion relation $k=\omega/c_s+ia\omega^2$ with $a=(4\eta/3+\xi)/(2\rho_0c_s^3)$, in agreement with the literature \cite{landau}. The fact that the wave vector is complex-valued when viscosity is present lacks the construction of a simple geometrical description for the perturbations. We wonder if other mathematical structures may prove to be more suitable for such purpose, such as Finsler or K\"ahler geometries \cite{finsler}, but this is the matter for future work.

Thus, if one insists in describing the small perturbations on the fluid motion geometrically, the above general procedure should be slightly modified when viscosities are present. This modification can be easily implemented through the Hadamard-Papapetrou method. Then, we shall assume that $\rho$ and $v_i$ are continuous functions through the wavefront $\Sigma_t(x_i)= \mbox{const.}$, for any given instant of time \(t\) (and, more generally, throughout spacetime), but with a possibly nonzero finite step in their derivatives across \(\Sigma_t\). For simplicity, we also consider an equation of state $p=p(\rho)$. The fundamental hypothesis here is that $\pi_{ij}$ is constructed by taking into account solely the background velocity of the fluid, without any contribution arising from the perturbations. This is acceptable since the viscosity coefficients are small in comparison with the background quantities as well as with the speed of sound in the medium. In other words, we assume that the background fluid moves with a continuous velocity, the spatial derivatives of which being taken to be continuous as well. Therefore, the step of the other quantities characterizing the fluid and its perturbations are
\begin{equation}
\begin{array}{ll}
\left[\rho_{,t}\right]_{\Sigma_t}=-\omega\tilde\rho, & \quad\left[\rho_{,i}\right]_{\Sigma_t}=\tilde\rho\,k_i, \\ [2ex]
\left[v_{i,t}\right]_{\Sigma_t}=-\omega \tilde v_{i}, & \quad\left[v_{i,j}\right]_{\Sigma_t}=\tilde{v_{i}} k_j,
\end{array}
\label{nr_hpd}
\end{equation}
where $\omega$ is the wave frequency and $k_i$ is the wave vector, which appear in virtue of the time and spatial derivatives, respectively. The arbitrary functions $\tilde\rho$ and $\tilde v_{i}$ describe the step of the corresponding quantities $\rho$ and $v_{i}$ through $\Sigma_t$. Note that the symbol
\begin{equation}
\left[f\right]_{\Sigma_t}(P)=\lim_{\delta\to0^+}[f(P_+)-f(P_-)]
\end{equation}
indicates how the step of an arbitrary function $f(x)$ through the borderless surface $\Sigma_t$ is evaluated, where $P\in\Sigma_t$ is arbitrary. For any \(\delta>0\), Let \(U_\delta(P)\) be a neighbourhood of \(P\) of radius \(\delta\). For sufficiently small \(\delta\), if follows that \(U_\delta(P)\) is splitted by \(\Sigma_t\) into three mutually disjoint components \(U_\delta(P)=U_\delta^+(P)\cup U_\delta^-(P)\cup U_\delta^o(P)\), where \(U_\delta^o(P)\subset\Sigma_t\) and \(U_\delta^\pm\) are chosen at opposite sides of \(\Sigma_t\) in a consistent manner ({\em i.e.}, such that, for any pair of points \(P,P'\in\Sigma_t\), we have \(U_\delta^+(P)\cap U_\delta^-(P')=\emptyset\)). Then, \(P_+\in U_\delta^+(P)\) and \(P_-\in U_\delta^-(P)\) are arbitrarily chosen.

Substitution of Eqs.~(\ref{soundbkg}) and (\ref{nr_hpd}) into the discontinuities of Eqs.~(\ref{nr_cont}) and (\ref{nr_euler}) yields
\begin{equation}
\omega\tilde\rho - \tilde\rho v_i k_i-\rho\tilde v_i k_i=0,
\label{nr_cont_hp}
\end{equation}
from the continuity equation, and
\begin{equation}
\omega\tilde\rho v_i + \omega\rho \tilde v_i - \tilde\rho c_s^2 k_i -\tilde\rho v_i v_j k_j -\rho \tilde v_i k_jv_j - \rho v_i\tilde v_j k_j =0,
\label{nr_euler_hp}
\end{equation}
from the Euler equation. If we isolate $\tilde\rho$ in Eq.\ (\ref{nr_cont_hp}) and substitute it into Eq.\ (\ref{nr_euler_hp}), we obtain an eigenvalue equation for $\tilde v_j$
\begin{equation}
\label{ev_eq_v}
\left(\Omega^2\delta_{ij}-c_s^2k_ik_j\right)\tilde v_j=0,
\end{equation}
where $\Omega:=\omega - k_j v_j$. The expression in parenthesis is called {\it acoustic Fresnel matrix} in analogy to the Fresnel equation obtained from the limit of geometrical optics \cite{nov_vis_volo}, that is $Z_{ij} = \Omega^2 \delta_{ij} - p' k_i k_j$. Non trivial solutions for this equation are obtained from $\det[Z_{ij}]=0$. This requirement yields
$$\Omega^4(\Omega^2-c_s^2k^2)=0,$$
which leads to the two possible expressions for the phase speed
\begin{equation}
\begin{array}{ll}
(v_{\rm{ph}})_1=\hat k_iv_i, &\qquad {\rm when}\quad \Omega=0,\\[2ex]
(v_{\rm{ph}})_2=\hat k_iv_i +c_s, &\qquad {\rm when}\quad \Omega=c_sk,
\end{array}
\end{equation}
where $\hat k_{i}=k_i/k$ and $k=\sqrt{k_ik_i}$. Note that the first possibility $(v_{\rm{ph}})_1$ corresponds solely to the drag effect due to the fluid motion. The amplitude of the discontinuities in this case --- the corresponding eigenvector of the acoustic Fresnel matrix with null eigenvalue --- is given by any vector lying on the two-dimensional space orthogonal to the wave vector $k_i$. In fact, this solution disappears if we use a reference frame that comoves with the fluid. Therefore, it has no physical interest and shall be neglected from our analysis from now on. The second possibility \((v_{\rm{ph}})_2\) contains the contribution from the speed of sound $c_s$ itself. We see that the viscosity of the background fluid does not alter the propagation of perturbations, as one could expect from the phenomenology in the limit of small velocities of the fluid.

On the other hand, we could be tempted to include dissipative terms in the above equations by relaxing conditions stated in Eq.\ (\ref{nr_hpd}) and allowing the first derivatives of $v_i$ to be continuous, but with its second derivatives presenting a step. In such case, Hadamard-Papapetrou method would lead only to algebraic restrictions to the viscosity coefficients $\eta$ and $\xi$, without any geometric description for the perturbations.

\section{Geometric acoustics for relativistic fluids} \label{II}
In this section, we show that it is possible to provide a geometrical description for the perturbations in the context of relativistic fluid mechanics, in spite of the unfruitful attempts concerning the aforementioned non-relativistic case. The basic difference lies in an extra term in the continuity equation due to the presence of viscosity. This extra term is linear in the anisotropic pressure and contains only first order derivatives of the fluid velocity. Thus, we shall see that the application of the Hadamard-Papapetrou method in this case yields an extra term in the effective metric.

In the relativistic version of fluid mechanics, the fluid itself is represented by an energy-momentum tensor of the form
\begin{equation}
T_{\mu\nu}= \rho V_{\mu}V_{\nu} - (p+\beta) h_{\mu\nu} +q_{\mu}V_{\nu} + q_{\nu}V_{\mu}+ \pi_{\mu\nu},
\end{equation}
where $\rho$ is the energy density, $p$ is the isotropic thermodynamical pressure, $\beta$ is the bulk viscosity, $q_{\mu}$ is the heat flux and $\pi_{\mu\nu}$ is the anisotropic pressure.%
\footnote{This means \(q_\nu V^\nu=0,\,\pi_{\mu\nu}=\pi_{\nu\mu},\,\pi_{\mu\nu}V^\nu=0,\,\pi^\nu{}_\nu=0\). }
All these components are relative to the normalized timelike four-vector $V^{\mu}$ (such that $V^{\mu}V_{\mu}=1$), and the tri-dimensional space orthogonal to \(V^\mu\) is defined by the spatial projector $h_{\mu\nu} = g_{\mu\nu} - V_{\mu} V_{\nu}$, where $g_{\mu\nu}$ is the background metric.

Conservation laws applied to such energy-momentum tensor, namely $T^{\mu\nu}{}_{;\nu}$=0, where the semi-colon symbol (\,$;$\,) means covariant derivative with respect to the background metric \(g_{\mu\nu}\), lead to the relativistic generalizations of Eqs.\ (\ref{nr_cont}) and (\ref{nr_euler}). Namely,
\begin{equation}
\label{relat_cont}
\dot\rho + (\rho +p+\beta)\,\theta+\dot q_{\mu}V^{\mu}  + q^{\nu}{}_{;\nu} - \pi^{\mu\nu}\sigma_{\mu\nu}=0,
\end{equation}
for the continuity equation, and
\begin{equation}
\label{relat_euler}
\begin{array}{l}
(\rho +p)\, a^{\mu} - p_{,\nu} h^{\mu\nu} -\beta_{,\nu} h^{\mu\nu}  + \dot q^{\alpha}h^{\mu}_{\alpha} +\theta q^{\mu} + \theta^{\mu\nu}q_{\nu} +\\[1ex]
\omega^{\mu\nu}q_{\nu} + h^{\mu}_{\alpha}\pi^{\alpha\nu}{}_{;\nu}=0,
\end{array}
\end{equation}
for the Euler equation. The kinematical quantities associated to the vector field $V^{\mu}$ are the expansion coefficient $\theta=V^{\mu}{}_{;\mu}$, the acceleration vector $a^{\mu}=V^{\mu}{}_{;\nu}V^{\nu}$, the expansion tensor $\theta_{\mu\nu}=\sigma_{\mu\nu}+\frac{1}{3}\theta h_{\mu\nu}$, the traceless shear tensor $\sigma_{\mu\nu} = \frac{1}{2}h_{\mu}^{\alpha}h_{\nu}^{\beta} (V_{\alpha;\beta}+V_{\beta;\alpha})-\frac13 \theta h_{\mu\nu}$ and the vorticity tensor $\omega_{\mu\nu} = \frac12 h_{\mu}^{\alpha}h_{\nu}^{\beta} (V_{\alpha;\beta}-V_{\beta;\alpha})$.

If the dissipative terms are determined and an equation of state is somehow provided, either through phenomenological relations or else by using dynamical equations, then Eqs.\ (\ref{relat_cont}) and (\ref{relat_euler}) completely describe the time evolution of the fluid.

\subsection{Effective acoustic metric for perfect fluids revisited}
\label{perfect}
In this section, we shall use the Hadamard-Papapetrou method to derive the effective acoustic metric for ideal barotropic fluids. Although the result is well known in the literature \cite{visser_molina}, the standard approach to get it from the dynamical equations for the perturbations is not as simple as the approach employed here. Thus, a few comments on this seem to be appropriate.

As we have seen before, in the limit of geometric acoustics, $\rho$ and $p$ as well as the comoving vector field $V^\mu$ are all continuous through $\Sigma_t(x_i)= \mbox{constant}$. The step in their derivatives are then determined by
\begin{equation}
\left[\rho_{,\mu}\right]_{\Sigma_t}=\tilde\rho\,k_\mu, \quad \left[p_{,\mu}\right]_{\Sigma_t}=\tilde p\,k_\mu, \quad \left[V_{\alpha,\mu}\right]_{\Sigma_t}=\tilde V_{\alpha}\,k_\mu
\label{h_k}
\end{equation}
where $k_\mu$ is the covariant wave 4-vector, the components of which are written as $k_\mu=(\omega,-\vec k)$.

In the case of barotropic fluids $p=p(\rho)$, where heat fluxes and viscosities are absent, the application of the Hadamard-Papapetrou method to the energy-momentum conservation laws leads to
\begin{equation}
\tilde\rho\,k_{\alpha} V^{\alpha} + (\rho+p)\,\tilde V^{\beta}k_{\beta}=0,
\end{equation}
and
\begin{equation}
- c_s^2\,\tilde\rho\, k_{\alpha} h^{\mu\alpha} + (\rho+p)h^{\mu\beta}\tilde V_{\beta}k_{\alpha}V^{\alpha}=0.
\end{equation}
The combination of these two equations yields an eigenvalue problem for $\tilde V^{\alpha}$ as
\begin{equation}
Z^{\mu}{}_{\alpha}\tilde V^{\alpha}=0,
\label{eigenV}
\end{equation}
where
\begin{equation}
Z^{\mu}{}_{\alpha}=\omega^2h^{\mu}_{\alpha}+c_s^2 h^{\mu\nu} h^\lambda_\alpha k_\lambda k_{\nu}
\label{Fresnel}
\end{equation}
is defined as the tensorial acoustic Fresnel matrix. The wave frequency is defined as $\omega=k_{\beta}V^{\beta}$. Similarly to the non-relativistic case, the existence of non trivial solutions of Eq.\ (\ref{eigenV}) requires
\begin{equation}
\mbox{\(\det_{(3)}\)}[Z^{\mu}{}_{\alpha}]=0,
\label{eqFresnel}
\end{equation}
where $\det_{(3)}[Z^{\mu}{}_{\nu}]$ means the determinant of the $3\times3$ matrix representation of \(Z^{\mu}{}_{\nu}\), since this matrix is orthogonal to \(V^\mu\). It is then immediate to obtain the phase velocity as
\begin{equation}
\label{ph_vel_id_flu}
v_{{\rm ph}}=\frac{\omega}{k}=c_s,
\end{equation}
where $k=\sqrt{-h^{\mu\nu}k_{\mu}k_{\nu}}$.

Equation (\ref{eqFresnel}) leads to $[(c_s^{-2}-1) V^{\mu}V^{\nu} + g^{\mu\nu}] k_\mu k_\nu=0$, from where we read the effective inverse metric tensor,
\begin{equation}
\hat g^{\mu\nu}=(c_s^{-2}-1)V^{\mu}V^{\nu}+g^{\mu\nu}.
\end{equation}
Notice that the perturbations propagate as null geodesics with respect to  \(\hat g^{\mu\nu}\).

Now we define the effective metric tensor $\hat g_{\mu\nu}$ by means of
$\hat g^{\mu\alpha}\hat g_{\alpha\nu}=\delta^{\mu}_\nu$, yielding
\begin{equation}
\hat g_{\mu\nu}=(c_s^2-1)V_{\mu}V_{\nu}+g_{\mu\nu}.
\end{equation}
This result coincides with the Gordon-like effective metric obtained from the eikonal approximation (see for instance Ref.\ \cite{visser_molina}). Since in our calculations the dynamical equations for the perturbations are not needed, because the dispersion relation is enough to our purposes, we then impose less requirements upon the fluid. For instance, $V^{\mu}$ is not required to be irrotational. The determinant of the effective metric is $\sqrt{-\hat g}=|c_s|\sqrt{-g}$, which means that the propagation of perturbations is well-defined as long as $g_{\mu\nu}$ keeps its Lorentzian signature. From the causal structure of $\hat g_{\mu\nu}$ it follows that any light-like vector in $\hat g_{\mu\nu}$ is a time-like vector in the background metric \(g_{\mu\nu}\), as \(c_s<1\) is always smaller than 1 (that is, the sound-cone of \(\hat g_{\mu\nu}\) lies inside the Minkowski light-cone).

\subsection{Effective acoustic metric for viscous fluids}
\label{viscous}
Let us assume that the fluid exhibits a non-zero anisotropic pressure described by $\pi_{\mu\nu}$. Contrary to the non-relativistic case, now $\pi_{\mu\nu}$ can also modify the continuity equation [see Eq.\ (\ref{relat_cont})]. We shall use this to show that $\pi_{\mu\nu}$ can contribute to the tensorial acoustic Fresnel matrix consistently with the non-relativistic case. To do so, we consider that \(\pi_{\mu\nu}\) is continuous through $\Sigma_t$ with its first derivatives being such that its divergence, when projected by $h_{\mu\nu}$ on the spatial hypersurfaces, are supposed to be continuous as well. This means that
\begin{equation}
[\pi_{\mu\nu}]_{\Sigma_t} = 0,\qquad [\pi_{\nu\alpha,\beta}]_{\Sigma_t}=\tilde{\pi}{}_{\nu\alpha}\,k_\beta,
\end{equation}
where the coefficients \(\tilde{\pi}{}_{\nu\alpha}\) constitute a symmetric traceless matrix orthogonal to the wave vector:
\begin{equation}
\tilde{\pi}{}_{\nu\alpha}=\tilde{\pi}{}_{\alpha\nu},\quad
g^{\nu\alpha}\tilde{\pi}{}_{\nu\alpha}=0,\quad
h^{\mu\nu}g^{\alpha\beta}\tilde{\pi}{}_{\nu\alpha}k_\beta=0.
\end{equation}
It turns out that \(\pi_{\alpha\beta}\) may present discontinuities of its projected derivatives along $V^{\mu}$, namely
\begin{equation}
\label{dis_pi}
\tilde{\pi}{}_{\nu\alpha}V^\alpha=-\pi_{\nu\alpha}\tilde V{}^\alpha,
\end{equation}
where the equality is obtained using a total derivative and the orthogonality condition between $\pi_{\mu\nu}$ and $V^{\alpha}$. From Eq.\ (\ref{h_k}), it follows that the right-hand side of Eq.\ (\ref{dis_pi}) has a non-vanishing step. This implies that the anisotropic pressure can modify the relation between the amplitudes of the steps by means of the continuity equation, although the Euler equation remains the same.

After some manipulation with these equations for the discontinuities into the equations for the fluid motion, Hadamard-Papapetrou method leads to an extra term correction for the acoustic Fresnel matrix Eq.~(\ref{Fresnel}) as
\begin{equation}
Z^{\mu}{}_{\alpha}=\omega^2 h^{\mu}_{\alpha}+c_s^2h^{\mu\beta}h^\lambda_\alpha k_\lambda k_{\beta}- \frac{c_s^2}{\zeta} h^{\mu\beta}k_{\beta}\pi^{\gamma}{}_{\alpha}k_{\gamma},
\end{equation}
with $\zeta=\rho+p$ being the relativistic enthalpy. Again, the vanishing of the 3-determinant of $[Z^{\mu}{}_{\alpha}]$ gives the dispersion relation, thus yielding the phase velocity
\begin{equation}
v_{{\rm ph}}^2=c_s^2\left(1+\zeta^{-1}\hat l_{\alpha}\hat l_{\beta}\pi^{\alpha\beta}\right),
\end{equation}
where $l_{\alpha}=h_{\alpha}^{\beta}k_{\beta}$ and $\hat l_{\mu}=l_{\mu}/l$ with $l=\sqrt{-l^{\mu}l_{\mu}}$. Notice that the anisotropic pressure can now modify the phase velocity of the perturbations, both increasing and decreasing their effective speed being possibilities in this case.

The inverse effective metric then takes the form
\begin{equation}
\hat g^{\mu\nu} = c_s^{-2} V^{\mu}V^{\nu} + h^{\mu\nu} - \zeta^{-1}\pi^{\mu\nu},
\end{equation}
and the corresponding effective acoustic metric reads
\begin{equation}
\label{effect_met}
\hat g_{\mu\nu} = AV_{\mu}V_{\nu} + Bh_{\mu\nu} + C\pi_{\mu\nu} + D\pi_{\mu\alpha}\pi^{\alpha}{}_{\nu},
\end{equation}
where the coefficients \(A,B,C,D\) are given by
\begin{eqnarray}
&&A=c_s^2,\qquad B=1+\frac{\det_{(3)} [\pi^{\mu}{}_{\nu}]}{\zeta^3\Upsilon},\\[1ex]
&&C=\frac{1}{\zeta\,\Upsilon},\qquad D=\frac{1}{\zeta^2\,\Upsilon},
\end{eqnarray}
with $\Upsilon = 1 - (1/2\zeta^2) \pi^{\mu}_{\nu} \pi^{\nu}_{\mu} - (1/\zeta^3)\det_{(3)}[\pi^{\mu}{}_{\nu}]$. In this case, the determinant of the effective metric \(\hat g_{\mu\nu}\) is $\sqrt{-\hat g}=\Upsilon^{-\frac{1}{2}}|c_s|\sqrt{-g}$. Then, the Lorentzian signature of the background metric is not enough to guarantee that the effective acoustic metric is Lorentzian as well, but it is also necessary to know the sign of $\Upsilon$ everywhere. If the viscosities are small, then $\Upsilon$ is positive. Recall that, for vanishing $\pi^{\mu}{}_{\nu}$, we recover the previous results.

For an arbitrary light-like vector $X^{\mu}$ in the acoustic metric \(\hat g_{\mu\nu}\), its norm in the background metric is given by
\begin{equation}
\label{norm_x}
\begin{array}{lcl}
g_{\mu\nu}X^{\mu}X^{\nu} &=& \left[1 - c_s^2 + \fracc{c_s^2\,\det_{(3)}[\pi^{\mu}_{\nu}]}{\zeta(\zeta^2 - \frac{1}{2} \pi^{\mu}_{\nu} \pi_{\mu}^{\nu})} \right](X^{\alpha}V_{\alpha})^2 +\\[2ex]
&&- \fracc{\zeta\pi^{\alpha}{}_{\beta} + \pi^{\alpha}{}_{\nu}\pi^{\nu}{}_{\beta}}{\zeta^2 - \frac{1}{2} \pi^{\mu}_{\nu} \pi_{\mu}^{\nu}}\,X^{\beta}X_{\alpha}.
\end{array}
\end{equation}
Note that the norm of $X^{\mu}$ with respect to the background metric $g_{\mu\nu}$ is not fixed {\em a priori}, but it depends on algebraic functions of $\pi_{\mu\nu}$ instead. Therefore, the anisotropic pressure can change the sound-cone on both ways (either by increasing or decreasing the effective sound speed) according to its magnitude in each spatial direction.

Without loss of generality, we can choose coordinates $x^{\mu}=(x^0,x^1,x^2,x^3)$ such that the matrix representation of the spatial part of $\pi^{\mu}{}_{\nu}$ is given by a linear combination of two $3\times3$ traceless diagonal matrix, i.e.,
\begin{equation}
\label{gen_pi}
\pi^{\mu}{}_{\nu}=\pi_1\,{\rm diag}(2,-1,-1) + \pi_2\,{\rm diag}(0,1,-1),
\end{equation}
where $\pi_1$ and $\pi_2$ are arbitrary functions. The first term is responsible for a symmetric deformation in the $(x^2,x^3)$-plane, while it gives an opposite contribution along the $x^1$-direction. The second term produces only a deformation in that plane, with opposite magnitudes along each axis. No simple results follow from substitution of this decomposition formula in Eq.~(\ref{norm_x}) in the general case. However, there are simple cases which can give us an idea about the role played by the anisotropic pressure in modifying the effective sound-cone.

For the sake of simplicity, we take $x^{\mu}$ as Cartesian coordinates and the background metric as the flat Minkowski spacetime written in this coordinate system. We also consider the comoving vector $V^{\mu}$ given by $\delta^{\mu}_{0}$. First, we study the case where $\pi_2=0$ in Eq.\ (\ref{gen_pi}). Then, we take an arbitrary vector $X^{\mu}$ and compute the modifications on each component of its spatial velocity caused by the presence of the anisotropic pressure through Eq.\ (\ref{norm_x}). Basically, this will give two different possibilities depending on either the velocity is orthogonal or parallel to the plane of symmetry determined by $\pi^{\mu}{}_{\nu}$:
\begin{equation}
\label{vel_case_sym}
v_{\perp}^2=\frac{c_s^2}{1-\frac{2\pi_1}{\zeta}},\qquad v_{||}^2=c_s^2\left(1+\frac{\pi_1}{\zeta}\right).
\end{equation}
In order to avoid issues concerning causality, the sound-cone must be inside the light-cone. This requires that the changes caused by the viscosity should be small and this is precisely the case, since the anisotropic pressure, in particular $\pi_1$, is microscopically determined by the inter-particle potential energy while the relativistic enthalpy contains such term and all other kinds of energy necessary to construct the system. Notwithstanding, we see that the anisotropic pressure squeezes the sound-cone, deforming it differently according to the direction of the wave propagation. For positive values of $\pi_1$, it increases the spatial velocities in directions parallel to the plane of symmetry, but it decreases the velocities along the perpendicular direction. Opposite results apply to the $\pi_1<0$ case.

The second case to analyze is $\pi_1=0$ in Eq.\ (\ref{gen_pi}). Both the background metric and the comoving vector field are assumed to be the same as in the previous case, using also Cartesian coordinates. Again, we consider an arbitrary vector $X^{\mu}$ and compute the modifications in its spatial velocities due to $\pi^{\mu}{}_{\nu}$. Since this case presents no symmetries, we have three different situations for the spatial velocity of $X^{\mu}$:
\begin{equation}
\label{vel_case_null}
v_0^2=c_s^2,\qquad v_{\pm}^2=c_s^2\left(1\pm\frac{\pi_2}{\zeta}\right).
\end{equation}
Notice that along the $x^1$-direction the spatial velocity is unchanged by the anisotropic pressure; in the complementary directions, however, it decreases along $x^2$ (as given by $v_{-}^2$) and increases along $x^3$ (given by $v_{+}^2$), as long as $\pi_2>0$. It should be emphasized that all these spatial velocities are actually indicating the inclination of the effective sound-cone along each direction. Since all of them are proportional to $c_s$, which is small compared to the speed of light, they are therefore less than unit, and the causality is safe.

For completeness, we should mention that there are simple phenomenological profiles for the anisotropic pressure which can be used to construct the analogue models. This is the case of the Newtonian fluids, which are characterized by an anisotropic pressure tensor linearly proportional to the shear tensor, with constant coefficient of proportionality (called dynamic viscosity coefficient). Namely, $\pi_{\mu\nu}=\xi \sigma^{(o)}{}_{\mu\nu}$, with $\xi={\rm const.}$; when $\xi$ may depend on the energy density, it is usually called a viscoelastic fluid. Finally, we have the Stokesian fluids where the anisotropic pressure is a function of the expansion tensor $\theta^{\mu}{}_{\nu}$ \cite{ericksen}. The most general expression of $\pi^{\mu}{}_{\nu}$ in this case is given by
\begin{equation}
\pi^{\mu}{}_{\nu} = \alpha_0 h^{\mu}{}_{\nu} + \alpha_1 \theta^{\mu}{}_{\nu} + \alpha_2 \theta^{\mu}{}_{\alpha} \theta^{\alpha}{}_{\nu},
\end{equation}
where $\alpha_0$, $\alpha_1$ and $\alpha_2$ are functions satisfying the constraint $9\alpha_0+3\alpha_1\theta+\alpha_2(3\sigma^2+\theta^2)=0$. Terms which involve the expansion tensor in orders higher than 2 can be rewritten in terms of lower orders terms in \(\theta^\mu{}_\nu\) due to the Cayley-Hamilton theorem.

\section{Analog black hole from a viscous fluid}\label{bh}
As mentioned before, the geometric description of analog models of gravity in fluid mechanics does not demand the complete dynamics of perturbations, their kinematics being enough. Thus, we shall present here the main steps to derive analog models for static and spherically symmetric solutions of general relativity from the effective acoustic metric given by Eq.~(\ref{effect_met}). Details for the case of a Schwarzschild black hole are given afterwards.

Consider the background metric as the one to describe Minkowski spacetime in spherical coordinates $(t,r,\theta,\phi)$, and the vector field $V^{\mu}=\delta^{\mu}_0$ comoving with the fluid. Suppose that, in this background, the anisotropic pressure has the form $\pi^{\mu}{}_{\nu}=\mbox{diag}(0,2\pi_1,-\pi_1,-\pi_1)$.
The non-vanishing components of the effective metric turn out to be, up to a conformal factor $1+\pi_1/\zeta$, given by
\begin{equation}
\begin{array}{ll}
\label{effc_met_stat}
\hat g_{00}=c_s^2\left(1+\fracc{\pi_1}{\zeta}\right),&
\hat g_{11}=-\fracc{1+\frac{\pi_1}{\zeta}}{1-\frac{2\pi_1}{\zeta}},\\[2ex]
\hat g_{22}=-r^2,&
\hat g_{33}=\hat g_{22}\sin^2\theta.
\end{array}
\end{equation}

In terms of coordinates $(x^0,R,\theta,\phi)$, this metric fits the standard form
\begin{equation}
\hat g_{\mu\nu} = f\,\delta^0_\mu\delta^0_\nu - \frac{\delta^1_\mu\delta^1_\nu}{f} - R^2(\delta^2_\mu\delta^2_\nu + \sin^2\theta\,\delta^3_\mu\delta^3_\nu),
\label{stat_sph}
\end{equation}
if the quantities \(c_s,\,\pi_1/\zeta,\,r\) satisfy the conditions
\begin{equation}
\label{match_eq_gen}
c_s^2=\frac{(2+f)f}{3}, \quad\frac{\pi_1}{\zeta}=\frac{1-f}{2+f},
\quad r=R,
\end{equation}
where $f=f(R)$ is a given radial function. Note that the metric given by Eq.\ (\ref{stat_sph}) includes all well-known static and spherically symmetric solutions of the theory of general relativity: Schwarzschild, Reissner-Nordstr\"om, de-Sitter and combinations thereof.

From the equations of motion, we obtain that the continuity equation is identically satisfied, since the variables which describe the fluid do not depend on time. The only non-trivial component of the Euler equation is the one corresponding to the radial coordinate, which leads to
\begin{equation}
-\frac{dp}{dr}+2\,\frac{d\pi_1}{dr}+6\,\frac{\pi_1}{r}=0.
\label{euler}
\end{equation}

In order to see how the anisotropic pressure may indeed contribute to the effective metric, we now analyze the case of an analog Schwarzschild black hole in details. The radial function is then $f(R)=1-R_s/R$ with $R_s>0$ as the Schwarzschild radius characterizing the event horizon. Thus, conditions (\ref{match_eq_gen}) become
\begin{equation}
\label{match_eq}
c_s^2=\left(1-\frac{R_s}{r}\right)\left(1-\frac{R_s}{3r}\right), \quad\frac{\pi_1}{\zeta}=\left(\frac{3\,r}{R_s}-1\right)^{-1}.
\end{equation}
Recall that the positivity of $c_s^2$ imposes that the analog model is valid only for $r>R_s$. Notwithstanding, the fluid equations are defined in Minkowski spacetime, so we will make some comments concerning its behavior for $r<R_s$. Therefore, the substitution of Eqs. (\ref{match_eq}) into the Euler equation (\ref{euler}) yields
\begin{equation}
-\frac{dp}{d\bar r}+2\frac{d}{d\bar r}\left(\frac{p+\rho}{3\bar r-1} \right)+6 \frac{p+\rho}{(3\bar r-1)\bar r}=0,
\label{luc1}
\end{equation}
where we define the dimensionless variable $\bar r=r/R_s$. On the other hand, we can rewrite the speed of sound as
\begin{equation}
\frac{dp}{d\bar r}=\frac{(\bar r-1)(3\bar r-1)}{3\bar r^2}\frac{d\rho}{d\bar r}.
\label{auxiliar}
\end{equation}

The derivative of Eq.\ (\ref{luc1}) with respect to $\bar r$ together with Eq.\ (\ref{auxiliar}) yields a simple second order differential equation for $\rho(\bar r)$, as follows
\begin{equation}
\frac{d^2\rho}{d\bar r^2} = h\left(\bar r\right)\frac{d\rho}{d\bar r},
\label{luc5}
\end{equation}
where
\begin{equation}
h(\bar r)= \fracc{-2(\bar r-\bar r_1)(\bar r-\bar r_2)(\bar r^2-a_1 \bar r+b_1)}{\bar r(\bar r-1/2)(\bar r-\bar r_3)(\bar r^2-a_2 \bar r +b_2)},
\end{equation}
with the constants having approximate values given by
\begin{equation}
\begin{array}{lcl}
\bar r_1=0.644781, & \bar r_2 = 2.39199, & a_1 = 0.379897, \\  b_1 = 0.0540316, & b_2=0.141752, & a_2 = 0.64848,
\end{array}
\end{equation}
and
\begin{equation}
\bar r_3 = 1+ \frac{\sqrt[3]{9 - \sqrt{17}} + \sqrt[3]{9 + \sqrt{17}}}{3} \approx2.35152 .
\end{equation}
\begin{figure}[t]
\begin{center}
\includegraphics[scale=0.32]{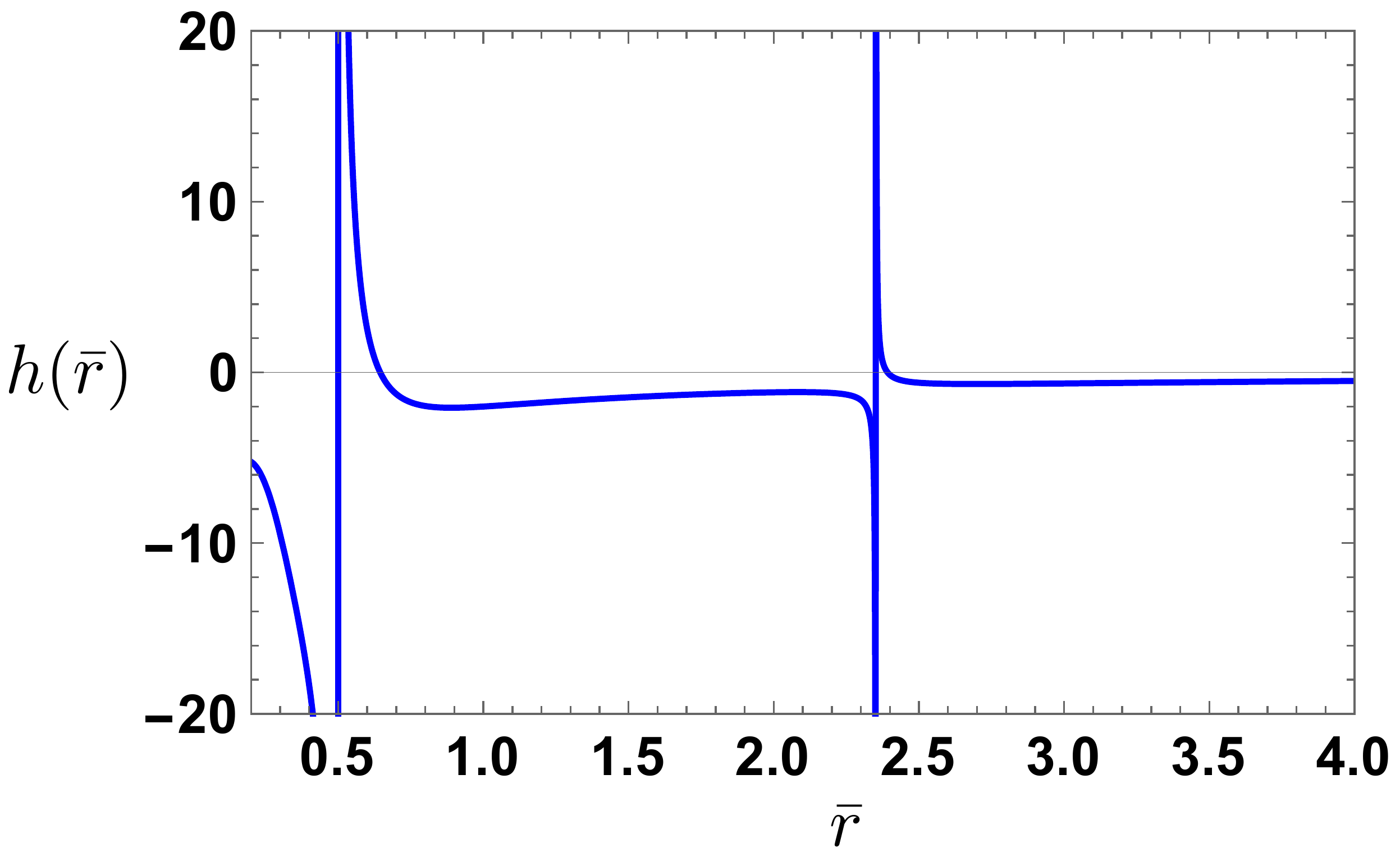}
\end{center}
\caption{Plot of $h(\bar r)$. Singularities occur at $\bar r=0$, $\bar r=1/2$ and $\bar r=\bar r_3$.}
\label{h}
\end{figure}
It should be noticed that the quadratic polynomials in the numerator and denominator are irreducible. Fig.\ \ref{h} depicts the behaviour of $h(\bar r)$.

In these coordinates, the Schwarzschild radius corresponds to $\bar r=1$. Note that $h(\bar r)$ is singular at $\bar r=0$, $\bar r=1/2$ and $\bar r=\bar r_3>1$. The initial conditions of Eq.\ (\ref{luc5}) are considered at $\bar r=1$ and, therefore, its general solution is
\begin{equation}
\rho(\bar r)=\rho(\bar r=1)+\rho'(\bar r=1)\int_{\bar r=1}^{\bar r}\exp\left[\int_{\bar r=1}^u h(v)dv\right]du.
\label{luc6}
\end{equation}

The last integrand in Eq.\ (\ref{luc6}) can be found explicitly as
\begin{eqnarray}
\exp\left[\int_{1}^{\bar r} h(v)dv\right] =  \left|\bar r-\bar r_3\right|^{\alpha_0} \left|6\bar r^2+a_3 \bar r +b_3 \right|^{\frac{\alpha_1}{12}}
\times\nonumber\\
\left|\frac{\bar r-\frac{1}{2}}{\bar r}\right|\exp\left[\frac{12 \alpha_2-a_3 \alpha_1}{6\sqrt{24b_3-a_3^2}} \arctan\left(\frac{12\bar r-a_3}{\sqrt{24b_3-a_3^2}}\right)\right]
\label{luc61}
\end{eqnarray}
where $\alpha_0\approx0.0358896$, $\alpha_1 \approx -12.2153$, $\alpha_2 \approx 3.41504$, $ a_3 = 6(\bar r_3-3)\approx-3.89088$ and $b_3 \approx 0.850514$ are new parameters. Hence, this expression is singular only at $\bar r=0$, and vanishes at both $\bar r=1/2$ and $\bar r=\bar r_3$.

Since the derivatives $p'(1)$ and $\rho'(1)$ are regular, the evaluation of Eqs.\ (\ref{luc1}) and (\ref{auxiliar}) at $\bar r=1$ provides the initial condition for the pressure as
\begin{equation}
p(1)=-\rho(1)-\frac{2}{3}\rho'(1).
\label{luc7}
\end{equation}
Since both pressure and energy density must be positive outside horizon in order to guarantee mechanical and thermodynamical stability of the fluid, Eq.\ (\ref{luc7}) implies that $\rho'(1)<0$. It is then straightforward to verify that all solutions of Eq.\ (\ref{luc6}) converge as $\bar r\rightarrow +\infty$.

Let $\bar r_0>1$ be the maximum radius of the fluid (including the limit case $\bar r_0=+\infty$). Thus,
\begin{equation}
\Delta \rho_0=\rho(\bar r_0)-\rho(1)=\rho'(1)I_\rho(\bar r_0)
\label{luc71}
\end{equation}
where
\begin{equation}
I_\rho(\bar r)=\int_{1}^{\bar r}\exp\left[\int_{1}^u h(v)dv\right] du.
\label{luc72}
\end{equation}

It follows from Eqs.\ (\ref{auxiliar}), (\ref{luc6}), (\ref{luc7}) and (\ref{luc71}) that
\begin{equation}
p(\bar r)=-\rho(1)+\frac{\Delta\rho_0}{I_\rho (\bar r_0)}\left[-\frac{2}{3}+I_p(\bar r)\right],
\label{luc73}
\end{equation}
where
\begin{equation}
I_p(\bar r)=\int_{1}^{\bar r}\frac{(u-1)(3u-1)}{3u^2}\exp\left[\int_{1}^u h(v)dv\right]du.
\label{luc74}
\end{equation}

Note that the condition (\ref{luc6}) imposes that
\begin{equation}
p(1)=-\frac{2}{3I_\rho(\bar r_0)}\rho(\bar r_0)+\left[\frac{2}{3I_\rho(\bar r_0)}-1 \right]\rho(1)\geq 0,
\end{equation}
which holds if, and only if
\begin{equation}
0\leq \rho(\bar r_0)\leq \left[1-\frac{3I_\rho (\bar r_0)}{2}\right]\rho(1).
\label{luc75}
\end{equation}
It is easy to see that inequality (\ref{luc75}) cannot hold for $\bar r_0>\bar r_4$, where \(\bar r_4=3.52635\) approximately. This means that this model does not allow an unbounded fluid, i.e., $\bar r_0$ arbitrarily large for a given set of the parameters. Since $\rho(\bar r)$ and $p(\bar r)$ are decreasing functions for $\bar r>1$, we can suppose that $\rho(\bar r_0)=0$. In this case, Eq.\ (\ref{luc73}) takes the form
\begin{equation}
\frac{p(\bar r)}{\rho(1)}=-1+\frac{1}{I_\rho (\bar r_0)}\left[\frac{2}{3}-I_p(\bar r)\right].
\label{luc76}
\end{equation}

\begin{figure}[ptb]
\begin{center}
\includegraphics[scale=0.32]{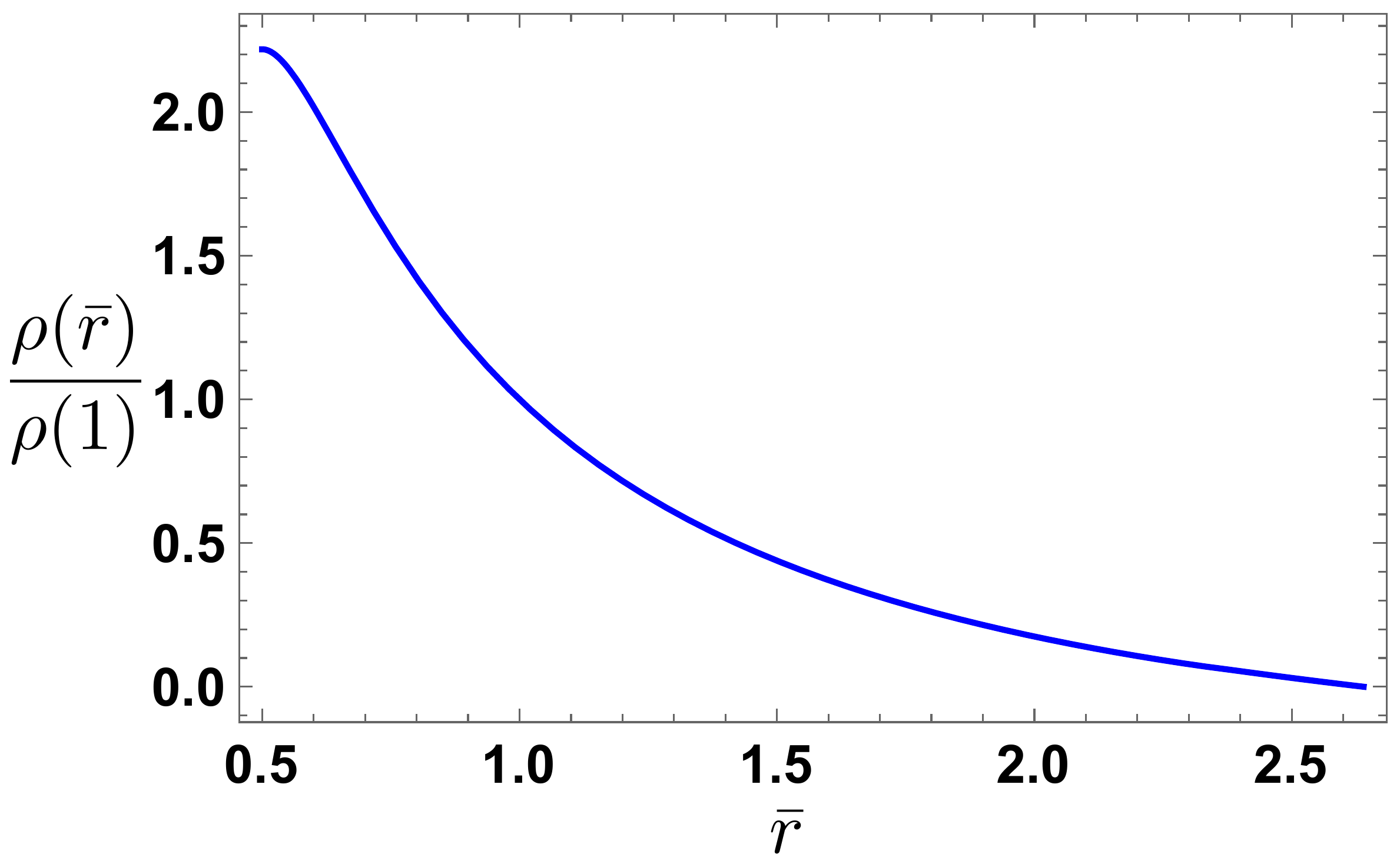}
\end{center}
\caption{Energy density as function of $\bar r$.}
\label{dens}
\end{figure}

\begin{figure}[ptb]
\begin{center}
\includegraphics[scale=0.32]{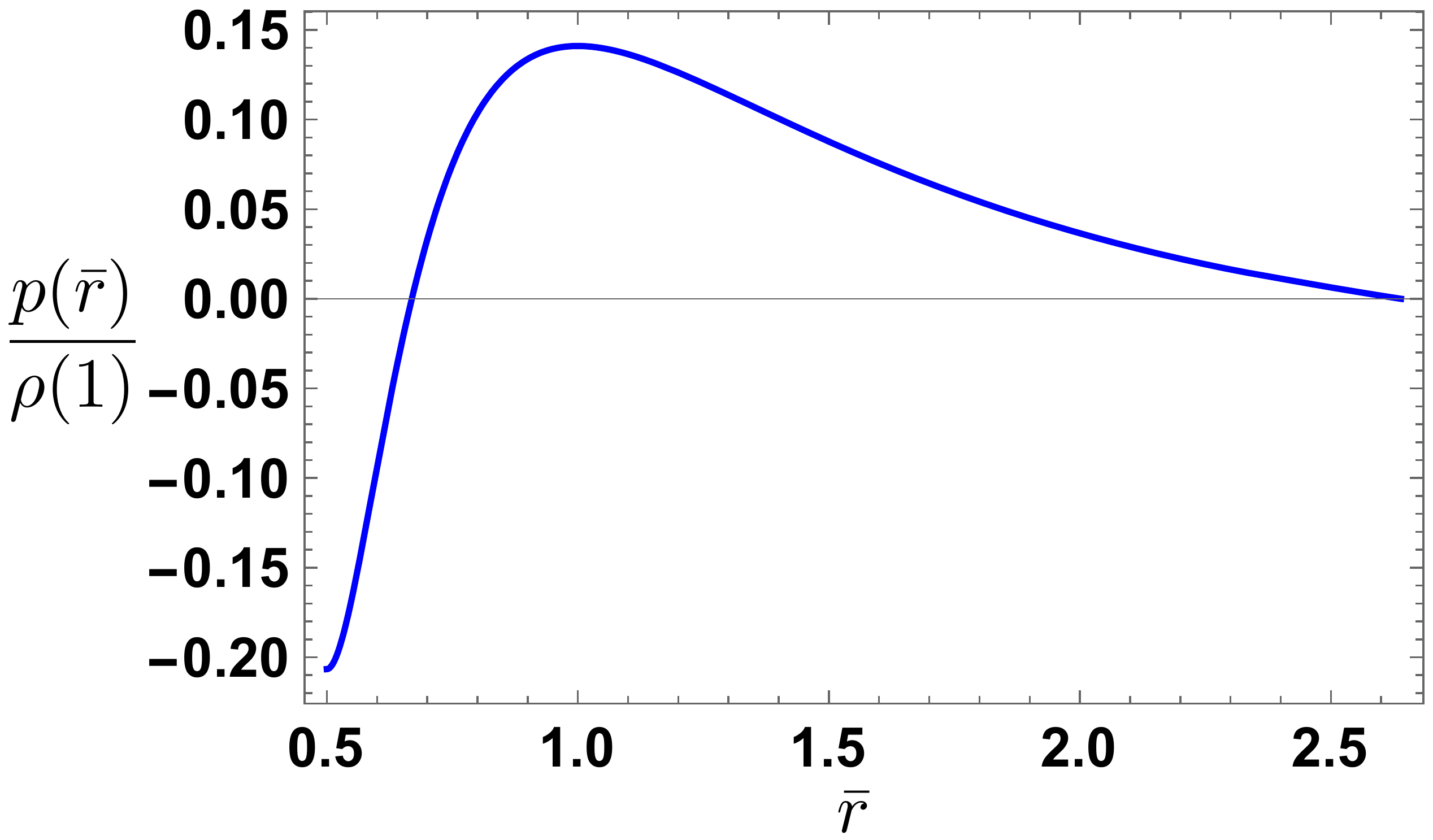}
\end{center}
\caption{Pressure as function of $\bar r$.}
\label{press}
\end{figure}

For any given choice of $\bar r_0$ in the interval $1<\bar r_0<\bar r_4$, Eq.\ (\ref{luc76}) provides the equilibrium pressure of the fluid with its surrounding medium to be $p_{ext}=p(\bar r_0)$.
\begin{figure}[t]
\begin{center}
\includegraphics[scale=0.32]{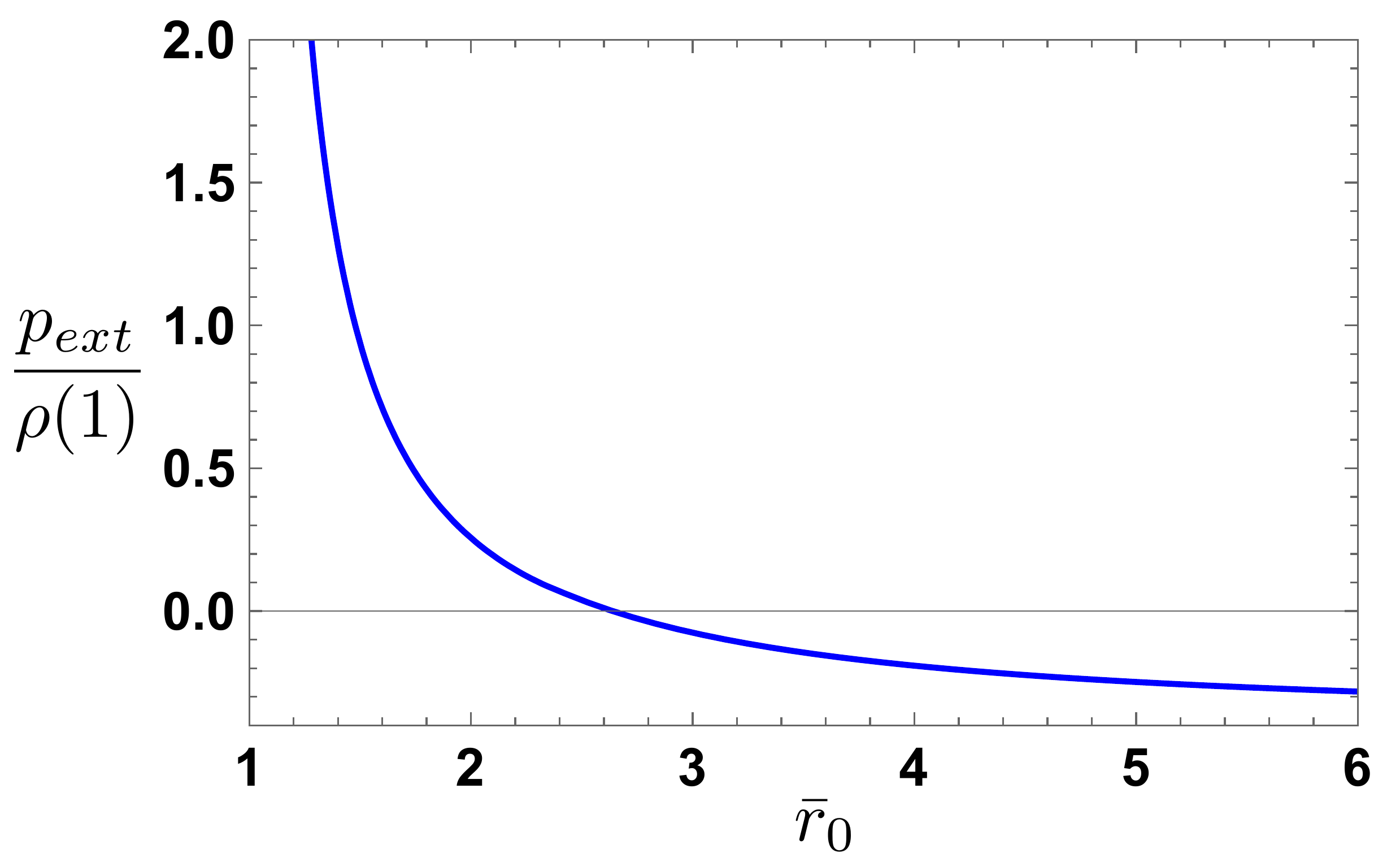}
\end{center}
\caption{The ratio $p_{ext}/\rho(1)$ as function of the maximum radius.}
\label{por}
\end{figure}
The behavior of $p_{ext}/\rho(1)$ is depicted in Fig.\ \ref{por}. It shows that the maximum radius cannot be bigger than $\bar r_{max}\approx2.63947<\bar r_4$, as anticipated, otherwise the pressure is negative from some $\bar r>1$. We stress that $\bar r_0=\bar r_{max}$ corresponds to $p_{ext}=0$. Hereafter, we analyze this special case, where there is no material medium surrounding the fluid. The corresponding behavior of the energy density and pressure, both normalized by $\rho(1)$, are depicted in Figs.\ \ref{dens} and \ref{press}, respectively.

\begin{figure}[ptb]
\begin{center}
\includegraphics[scale=0.32]{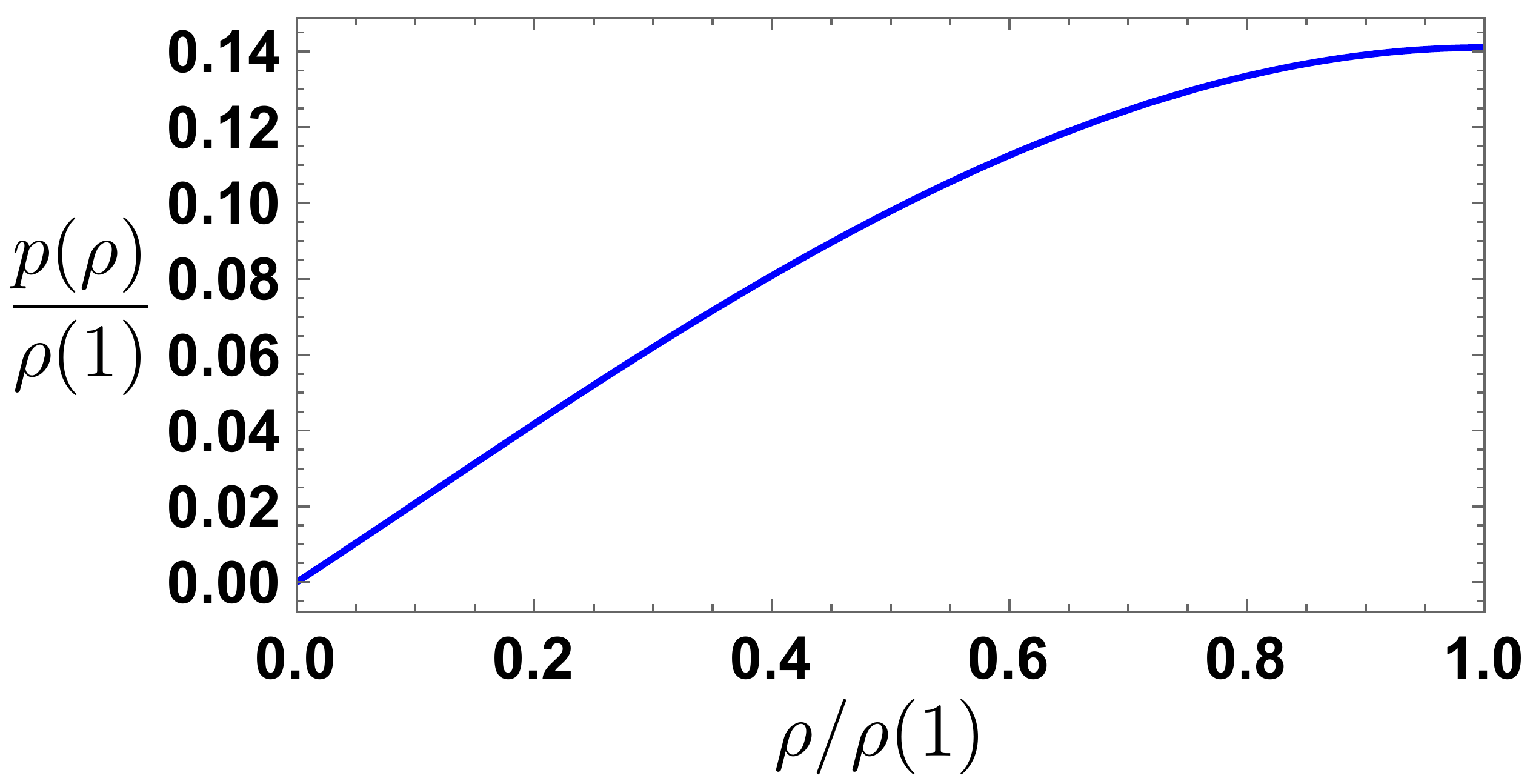}
\end{center}
\caption{Equation of state $p=p(\rho)$ profile.}
\label{pressrho}
\end{figure}

\begin{figure}[ptb]
\begin{center}
\includegraphics[scale=0.32]{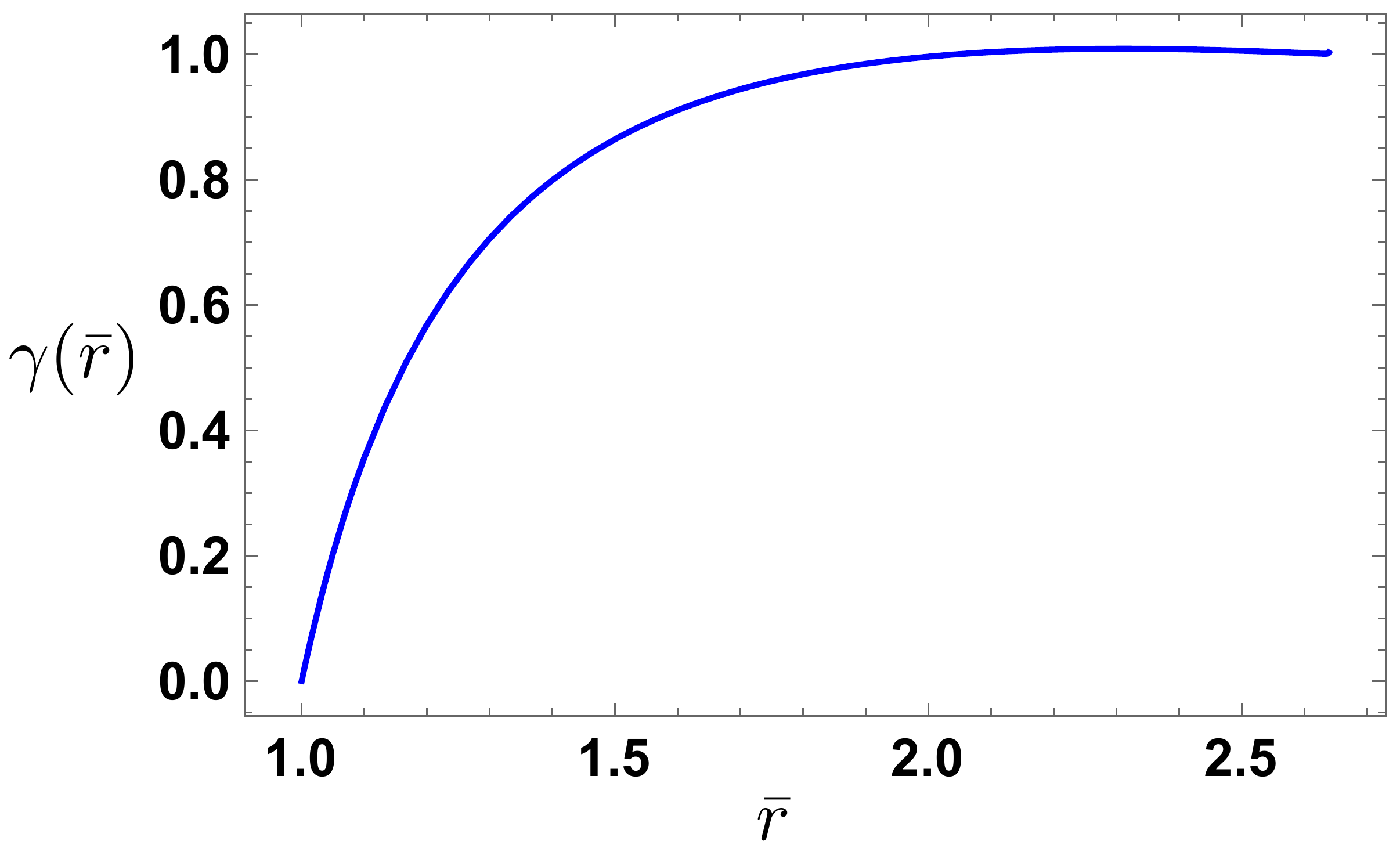}
\end{center}
\caption{Exponent $\gamma$ as function of $\bar{r}$.}
\label{exps}
\end{figure}

For the vacuum case, both $p(\bar r)$ and $\rho(\bar r)$ are completely determined by $\rho(1)$, and any two given solutions differ from each other only by a multiplicative constant. More precisely, $\rho(\bar r)=\rho(1)f_1(\bar r)$, $p(\bar r)=\rho(1)f_2(\bar r)$, where $f_1(\bar r)$ and $f_2(\bar r)$ are both decreasing functions in the range $1<\bar r<\bar r_{max}$. These two functions are also free of parameters. So, the equation of state of this fluid is determined by
\begin{equation}
p(\rho)=\rho(1) f_2 \circ f_1^{-1}\left(\frac{\rho}{\rho(1)}\right),
\end{equation}
which is illustrated in Fig.\ \ref{pressrho}. This equation of state depends on the choice of the energy  density at $\bar r=1$, but any two such functions are all equal up to a rescaling. Obviously, $p(\rho)$ is not a power law. However, around each radius $\bar r$, we can find the most suitable exponent $\gamma(\bar r)$ in order that their relation is described by
\begin{equation}
p=A \rho^\gamma.
\end{equation}
Such exponent \(\gamma\) can be calculated as
\begin{equation}
\gamma= \frac{\rho}{p} \frac{dp}{d\rho} =\frac{(\bar r-1)(3\bar r-1)}{3\bar r^2} \frac{f_1(\bar r)}{f_2(\bar r)}.
\end{equation}
It should be noticed that this expression is independent of $\rho(1)$, so the power law around each radius is completely determined. As it is apparent in Fig.\ \ref{exps}, the exponent runs from $\gamma=0$ (at $\bar r=1$) to $\gamma=1$ (at $\bar r=\bar r_{max}$).

\section{Concluding Remarks}
\label{V}
In this manuscript the propagation of acoustic perturbations in relativistic isentropic fluids was investigated. The effective metric associated with the propagation of these perturbations was formally obtained. Technicalities of the commonly used approach, which makes the treatment of viscous fluids a difficult task, were here circumvented by making use of the Hadamard-Papapetrou formalism. A natural proposal for the effective geometry experienced by the perturbations of a viscous fluid emerges as a result, which in turn makes apparent the importance of its relativistic description. Generalization of the results to the case of more general dissipative systems is expected to be but a simple exercise. As far as kinematical aspects are considered, analog models for general relativity can be obtained in terms of the physical properties of fluids. As an application, an acoustic black hole solution based on viscous fluid was discussed.

Analog models presenting event horizons have been investigated in several branches of physics. Perhaps the main point in studying such  special solutions is the possibility of digging a better understanding of Hawking radiation emission, which is a kind of process linked to the existence of an event horizon. In the same way as a gravitational black hole is expected to emit such thermal radiation, a corresponding acoustic radiation is expected to be produced in hydrodynamical systems whenever an analog event horizon is present. There is no real expectation that this really tiny effect in self-gravitating astrophysical systems will eventually be measured. Fortunately, this is not the case in the realm of analog models. In fact, measurement of this effect was quite recently claimed \cite{steinhauer2014,steinhauer2016} to be performed in a Bose-Einstein condensate system exhibiting an acoustic event horizon. This is an important achievement that, upon confirmation by other experimental groups, may lead to new insights in semi-classical gravity, as well as in quantum mechanics of hydrodynamic systems.

Another important issue related to the study of condensed matter analog models of gravitational black holes is the possibility of devising new devices with the property of being perfect energy absorbers. For instance, if a system is tailored in such way that energy is only transported by means of waves, then the existence of an analog black hole would be enough the guarantee that all carried energy would be trapped inside the event horizon. This issue deserves further examination.

\acknowledgments
The authors would like to thank the participants of {\it GIFT and PMAT Seminars} for comments on the manuscript. VADL acknowledge {\it Conselho Nacional de Desenvolvimento Cient\'{\i}fico e Tecnol\'ogico} (CNPq, grant 302248/2015-3) for partial support.

\end{document}